\documentclass[
aps,%
12pt,%
final,%
oneside,%
nobibnotes,%
nofootinbib,% 
superscriptaddress,%
centertags,
floatfix,
secnumarabic]%
{revtex4}%

\usepackage[utf8]{inputenc}
\usepackage{xcolor}
\usepackage[english]{babel}
\usepackage{amsmath}
\usepackage{amssymb}
\usepackage{graphicx}
\usepackage[hidelinks]{hyperref}

\begin{document}

\title{Pentaquarks with a hidden flavour and the hypothesis of the intrinsic charm  of the proton.}

\author{A. V. Berezhnoy}
\email{Alexander.Berezhnoy@cern.ch}
\affiliation{SINP MSU, Moscow,  119991, Russia}

\author{A. S. Gerasimov}
\email{Anton.Gerasimov@ihep.ru}
\affiliation{NRC "Kurchatov Institute", IHEP, Protvino, Moscow Region, 142281, Russia}

\newcommand{\MeV}{\,\text{MeV}}

\begin{abstract}
Using the experimental data on pentaquarks with a hidden charm, new  upper limits on the  intrinsic charm in the proton are obtained, possible shapes of heavy quark distributions  inside the proton are discussed.
\end{abstract}
  
\maketitle
\section{Introduction}
The hypothesis of the intrinsic charm in the proton was proposed quite a lot time ago to explain disagreement between early experiments on the charmed particles production and QCD predictions~\cite{Brodsky:1980pb}. However, as new data has been accumulated and a  calculation technique has been improved, it became clear that the assumption of intrinsic charm could not find any univocal confirmation.  Nevertheless, due to the elegance of this model and due to the occasional difficulties in describing charm events, this hypothesis attracts the attention of researchers for decades.  Now we expect a new wave of interest to the discussed model  because of the recent observation of three resonances in the $J/\psi \, p$ spectrum by the LHCb experiment\cite{Aaij:2019vzc}
% \cite{Aaij:2015tga,Aaij:2016phn} 
(see also the review ``Pentaquarks'' in the 2019 update of \cite{Tanabashi:2018oca}):

\begin{align}
    P_c(4312)^+ \; & (M=4311.9 \pm 0.7^{+6.8}_{-0.6} \MeV,\; \Gamma = 9.8 \pm 2.7^{+3.7}_{-4.5} \MeV), \nonumber \\
    P_c(4440)^+ \; & (M=4440.3 \pm 1.3^{+4.1}_{-4.7} \MeV,\; \Gamma = 20.6 \pm 4.9^{+8.7}_{-10.1} \MeV),  \label{eq:Pc}\\
    P_c(4457)^+ \; & (M=4457.3  \pm 0.6^{+4.1}_{-1.7} \MeV,\; \Gamma = 6.4  \pm 2.0^{+5.7}_{-1.9} \MeV )\nonumber. 
\end{align}

It is clear that states that strongly decays into $J/\psi$-meson and proton can be interpreted as $uudc\bar c$ pentaquarks. Moreover, it is possible that at least one of these states has the same quantum numbers as the proton ~\cite{Wang:2019got}, \cite{Zhu:2019iwm}, which means that the assumption of mixing such a pentaquark and a proton is quite reasonable. This assumption opens up a new view at the old problem  of intrinsic charm: now we can discuss the presence in proton of an admixture of a really existing baryon containing charm valence quarks.

It should be noted that initially the LHCb experiment announced the observation of a broad resonance $P_c(4380)$ and a narrow resonance $P_c(4450)$~\cite{Aaij:2015tga, Aaij:2016phn}. However, the latest research~\cite{Aaij:2019vzc}  showed, that $P_c(4550)$  splits into two narrow resonances: $P_c(4440)$ and $P_c(4457)$, and a  narrow resonance $P_c(4312)$ appeared near  $P_c(4380)$.  As a result,  the proof of the existence of $P_c(4380)$ became less convincing (\cite{Tanabashi:2018oca}). Therefore, in our work we discuss only three narrow resonances~\eqref{eq:Pc}. However, the possibility of the presence of $P_c(4380)$ cannot be completely discounted, since the new analysis is weakly sensitive to broad resonances.

In this work we study the possible distributions of $c$-quarks in the proton and estimate an upper limit on the intrinsic charm contribution within the nonperturbative model describing the mixing of a proton and a pentaquark based on \cite{Kuti:1971ph}.  The results are compared with the prediction of the intrinsic charm model  \cite{Brodsky:1980pb}.

\section{Distributions of heavy quarks in the proton in the framework of various models of intrinsic flavour}

In the article \cite{Brodsky:1980pb}, where the hypothesis of intrinsic charm was first discussed, it is assumed that the proton wave function can be represented as $|p\rangle = A_0 |uud\rangle + A_1 |uudc\bar c \rangle + \dots$, where $|A_1|^2$ --- nonzero probability of the existence of intrinsic $c\bar{c}$ -pairs. To estimate this probability, the ``old'' perturbation theory is used:

\begin{equation}
  G=|A_1|^2=\left|\frac{\langle u_1u_2 d c\bar c |M| u_1u_2d \rangle}{E_{u_1u_2dc\bar c}-E_{u_1u_2d}}\right|^2.
\end{equation}
 From the above expression in the system of infinite momentum and under the assumption that $\langle u_1u_2 d c\bar c |M| u_1u_2d \rangle=const$, one obtains, that
\begin{equation}
  G(x_{u_1},x_{u_2},x_{d},x_{c},x_{\bar c})\sim \left(M_p^2-\sum_{i=u_1,u_2,d,c,\bar c}\frac{m^2_{\perp i}}{x_i}\right)^{-2},
\end{equation}
which under assumption of very heavy charmed quarks ($m_c^2 \gg M_P^2,m_u^2,m_d^2$) leads to the following equation:
\begin{equation}
  G(x_{u_1},x_{u_2},x_{d},x_{c},x_{\bar c})\sim \frac{x^2_{c}x^2_{\bar c}}{(x_c+x_{\bar c})^2}\delta \left(1-\sum_{i=u_1,u_2,d,c,\bar c} x_i \right).
  \label{eq:ic_unint_dist}
\end{equation}

After the integration of (\ref{eq:ic_unint_dist}) one obtains a  distribution of single  $c$-quark in  the proton~\cite{Brodsky:1980pb}:
\begin{align}
    G_c(x_c) \sim x_c^2 \left[ (1-x_c)(1+10x_c+x_c^2)-6x_c(1+x_c)\ln \frac{1}{x_c} \right].
    \label{eq:brod_model}
    \end{align}
 The  normalization of  distribution (\ref{eq:brod_model}), which obviously determines the probability of finding a charmed quark in a proton, is not theoretically calculated in \cite{Brodsky:1980pb}. In a recent  theoretical study~\cite{Bednyakov:2017vck}  a restriction on the intrinsic charm of $1.93\% $ was obtained using the data of the ALTAS experiment. It should also be noted that there are studies that give greater limitations: in \cite{Mikhasenko:2012km} was obtained an upper limit of $1\%$ using the ratio of $\Lambda_ {QCD}$ to the difference in the energies of the pentaquark and proton, in  \cite{Litvine:1999sv} was obtained even  stricter restriction: $10^{-5} $.

Let us remind,  that equation (\ref{eq:brod_model})   is obtained in \cite{Brodsky:1980pb} under the assumption  of the very heavy charm  quark. This automatically leads to the idea  of not only intrinsic charm, but also intrinsic beauty. This is why, in the following sections we will discuss  this problem a bit. It is clear that the approach~\cite{Brodsky:1980pb}  predicts the same shapes of distribution of $c$ and $b$ quarks in a proton. On the contrary, as we will show later, in the model of proton and pentaquark mixing, these distributions must be different. 
Of couse we should not expect a sizible contribution of the internal beauty\footnote{See, for example, \cite{Polyakov:1998rb}, where it was predicted, that the internal beauty contribution in comparison with the internal charm contribution is suppressed as   $m_c^2/m_b^2$.}, however this problem is very interesting from the theoretical point of view.

To obtain distributed quarks in hadrons, there is a nonperturbative Kuti-Weisskopf model \cite{Kuti:1971ph},  which has successfully proven itself in the calculation of structure functions for small $q^2$. According to this model, the probability of detecting $n$ partons in the hadron, of which $m$ are valence, is determined by the expression:
\begin{align}
    G(x_1, \dots, x_n)\sim  \prod_{i=1}^m x_i^{1-\alpha_i} \quad \prod_{i=1}^n\frac{dx_i }{x_i} \quad \delta\left( 1 - \sum_{i=1}^n x_i \right).
\label{kuti_int}
\end{align}

In the discussed model the probability for sea parton is proportional to the phase volume $\frac{dx_i}{x_i}$. For valence quarks there is an additional factor $x_i^{1-\alpha_i}$, where the parameter $\alpha_i$ is related to the intersection of the Regge trajectory.

In the described approach, the pentaquark differs from the ``ordinary'' hadrons only in the presence of five valence quarks. Moreover, the value of the parameter $\alpha_c$ for valence $c$-quarks is known and it is equal to $-2.2$ \cite{Kartvelishvili:1985ac, Gershtein:2007nj}. As for the remaining values of $\alpha_i$, they are known from the Regge phenomenology:
$\alpha_u=\alpha_d=\alpha_q=1/2$  and $\alpha_s=0$.

Following \cite{Kuti:1971ph}, one can obtain the following distributions for the $uudc\bar{c}$ pentaquark (see~\cite{Mikhasenko:2012km}):
\begin{equation}
    G_q^{P_c}(x_q) \sim   x^{-\alpha_q} (1-x)^{-1+\gamma+2(1-\alpha_q)+2(1-\alpha_c)} 
\label{eq:Pc_lq_distr}    
\end{equation}
 for light valence quarks;
\begin{equation}
    G_c^{P_c}(x_c) \sim x^{-\alpha_c} (1-x)^{-1+\gamma+3(1-\alpha_q)+(1-\alpha_c)}
\label{eq:Pc_c_distr}    
\end{equation}
 for valence $c$ and $\bar c$ quarks. Because quarks are valence, the distributions (\ref{eq:Pc_lq_distr}) and (\ref{eq:Pc_c_distr}) are normalized to $1$.

Let us remind, that according to \cite{Kuti:1971ph} the distributions  of valence quarks  in proton have the following shape:
\begin{align}
    G^{p}_q (x_q)\sim x_q^{-\alpha_q}(1-x_q)^{-1+\gamma +2(1-\alpha_q)}.
\end{align}

For the proton  $\gamma=3$, and we will use the same value  to calculate the distribution of quarks in the pentaquark,  as it was done in~\cite{Mikhasenko:2012km}.

\begin{figure}
\begin{minipage}[h]{0.49\linewidth}
\center{\includegraphics[width=0.9\textwidth]{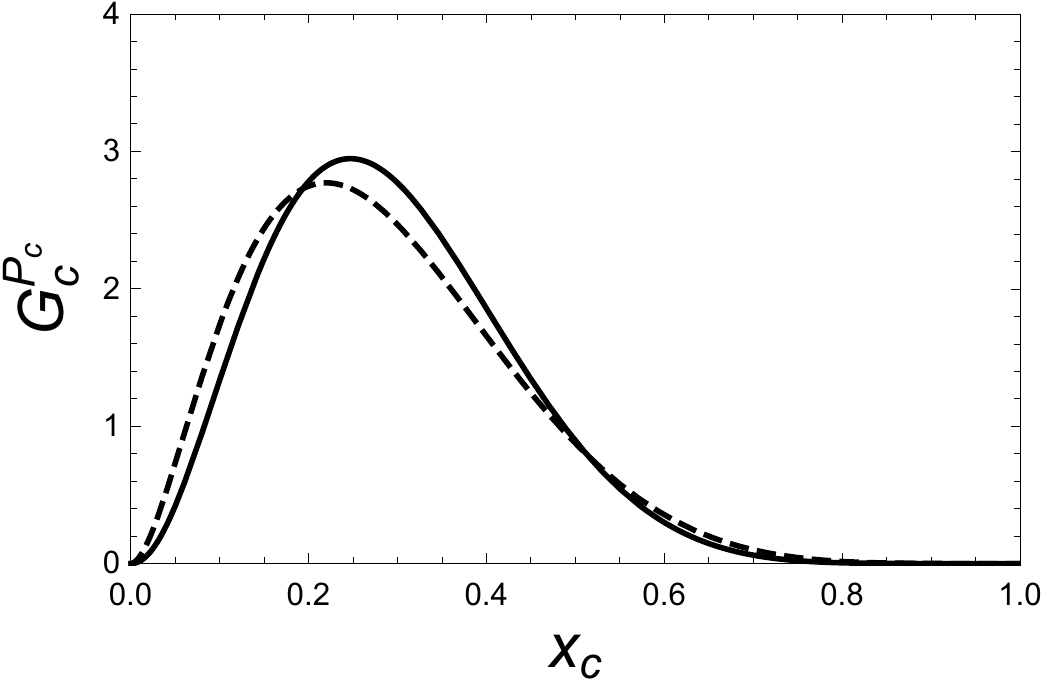}\\ (a)}
\end{minipage}
\begin{minipage}[h]{0.49\linewidth}
\center{\includegraphics[width=0.9\textwidth]{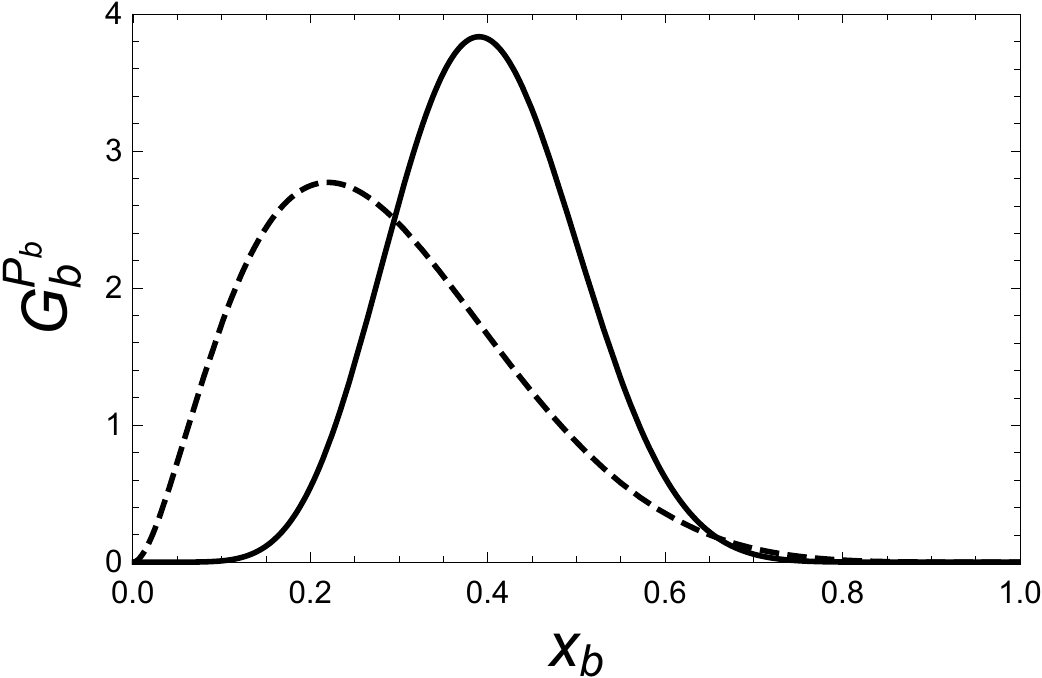}\\ (b)}
\end{minipage}
\caption{ (a) Distributions of $c$-quark in the proton according to the model \cite{Brodsky:1980pb} (dashed curve) and in $uudc\bar c$ pentaquark according to the model \cite{Kuti:1971ph} (solid curve) (see also~\cite{Mikhasenko:2012km}); (b) the analogous distributions for the $b$ quark case.}

\label{pic_cc}
\end{figure}

Comparing the distributions of charmed quarks in a proton obtained within the intrinsic charm model (\ref{eq:brod_model})  in a pentaquark (\ref{eq:Pc_c_distr}), it is notable that they have very similar shapes (see Fig.\ref{pic_cc}(a)). However, using the example of the hypothetical  $uudb\bar{b}$ pentaquark, it can be shown that this is just a coincidence. 

Indeed, the distribution of the valence $b$-quark in such a pentaquark can be obtained from (\ref{eq:Pc_c_distr}) with replacement of parameter $\alpha_c=-2.2$ to $\alpha_b=-8$ \cite{Kartvelishvili:1977pi}:
 \begin{equation}
    G_b^{P_b}(x_b)\sim  x^{-\alpha_b} (1-x)^{-1+\gamma+3(1-\alpha_q)+(1-\alpha_b)},
\end{equation}
which significantly changes its shape. Herewith, as we noted earlier, the distribution for the ``intrinsic beauty'' in the proton is exactly the same as for the intrinsic charm. In view of these circumstances, the difference in the  distribution shapes becomes very noticeable (see Fig.\ref{pic_cc}(b)), which means the similarities between (\ref{eq:brod_model}) and (\ref{eq:Pc_c_distr}) cannot be considered as a regularity.
 
It worth to note, that within the discussed model it is possible to obtain the strange quark distribution for a pentaquark with a hidden strangeness. Due to the fact that $\alpha_s=0$  this distribution have will the simplest form: 
 \begin{equation}
    G_s^{P_s}(x_s)\sim   (1-x_s)^{4.5}.
\end{equation}
  
Nevertheless, for today  pentaquarks with  a hidden strangeness were not observed, and  the discussion of these particles is purely theoretical, as well as the discussion of pentaquarks with a hidden beauty. It should also be noted that, unfortunately, the presence of internal strangeness, as well as the presence
internal charm, does not find unambiguous experimental confirmation (see the theoretical review \cite{Brodsky:2016fyh}, where  the experimental data of the HERMES experiment \cite{Airapetian:2013zaw} are discussed).
 
\section{Pentaquark  admixture of the proton}

Despite that the similarity in the distributions of $c$ quarks in the proton within the intrinsic charm model~\cite{Brodsky:1980pb} and in the pentaquark within the Kuti-Weisskopf model~\cite{Kuti:1971ph} is a coincidence, it prompts to the idea of a possible mixing of a proton and one of the recently discovered pentaquarks (\ref{eq:Pc}). Further, on the assumption that one of these pentaquarks has quantum numbers of a proton, one can estimate probability of such mixing and, therefore, the probability of detection of the pentaquark component in the proton.

The mixed states of proton and pentaquark $ X_1|p\rangle + X_2| P_c\rangle$ are defined with help of mixing matrix:

\begin{align}
    \begin{pmatrix}
        M_p& V\\
        V& M_{P_c}
    \end{pmatrix}
    \Bigg|
    \begin{matrix}
        X_1  \\
        X_2 
    \end{matrix} 
    \Bigg\rangle
    = \lambda
    \Bigg|\begin{matrix}
        X_1  \\
        X_2
    \end{matrix}\Bigg\rangle,  
\end{align}
  where $M_p$ and $M_{P_c}$ are masses of proton and pentaquark, and  $V$ is operator connected with their interaction.

Assuming the smallness of $V$ with respect to the mass difference $M_{P_c}-M_p$, we can find mixing matrix eigenvalues:
\begin{align}
    \lambda_1&=M_p + \frac{V^2}{M_{P_c}-M_p}, \nonumber \\ 
    \lambda_2&= M_{P_c} - \frac{V^2}{M_{P_c}-M_p},
\end{align}
Herewith the state of the proton with an admixture of pentaquark corresponds to the first eigenvalue
\begin{equation}
    |p\rangle^{\prime}= |p\rangle + \varepsilon |P_c\rangle,
\end{equation}
where the small parameter 
\begin{equation}
\varepsilon=\frac{V}{M_{P_c}-M_p},
\end{equation}
and the state of pentaquark with an admixture of proton corresponds to the second eigenvalue
\begin{equation}
     |P_c\rangle^{\prime}= |P_c\rangle + \varepsilon |p\rangle .
\end{equation}

The operator $V$ is unknown, but it can be assumed that it is bounded in value by the full width of  pentaquark
\begin{equation}
    V \lesssim\Gamma_{P_{c}}
\end{equation}
and therefore
\begin{equation}
    \varepsilon \lesssim \frac{\Gamma_{P_{c}}}{M_{P_c}-M_p}.
\end{equation}

Here we are forced to note that this assumption  has no strict justification and is only a guess. Our qualitative considerations are that the pentaquark mainly decays into a proton and $J/\psi$, which means that the wider the pentaquark, the greater the value of the operator of the transition to the proton. A more reasonable estimate of the value of $V$ will be the goal of our next study.

Which of the observed pentaquarks~(\ref{eq:Pc}) has the quantum numbers of a proton is not exactly known, but most likely this is the state with the smallest mass, since it is the main candidate for the ground state. We will use its width (9.8 MeV) to estimate the value of the admixture:

\begin{equation}
    \varepsilon \lesssim \frac{9.8 \mbox{ MeV}}{4312 \mbox{ MeV}-938 \mbox{ MeV}}  \approx 3 \cdot 10^{-3}.
\end{equation}

Thus, the probability of the presence of a five-quark state in a proton is  $\varepsilon^2 \lesssim  10^{-5}$.

It is worth to note,  that a very similar problem of mixing of light baryons with light pentaquarks  was presented in a recent article \cite{Xu:2020ppr}.

\section{Summary}

The intrinsic charm model \cite{Brodsky:1980pb} allows to determine the shape of  $c$ quark distribution in a high momentum system, but does not predict the normalization. On the contrary, the model of a proton and a pentaquark mixing allows not only to predict the distribution shape, but also to estimate the admixture of $c$ quarks in the proton. The  shape of $c$ quark distribution is predicted in the framework of the Kuti-Weisskopf \cite{Kuti:1971ph} approach applied to the pentaquark, and the admixture value is estimated using the masses and the widths of recently discovered LHCb pentaquarks with hidden charm~\cite{Aaij:2015tga, Aaij:2016phn}.

It is worth to note, that such different approaches lead to very similar $c$ quark distributions in a proton. However, most likely, this is a coincidence, as it  was demonstrated by example of  the $b$ quark  distributions  predicted in the framework of the same two models.

As an upper limit of  the admixture value $\varepsilon$ of the $c$ quark in the proton $|p^\prime \rangle = |p\rangle + \varepsilon |P_c\rangle$ we used the value of  ratio $\Gamma_{P_{c}}/(M_{P_c}-M_p)\approx 3 \cdot 10^{-3}$, where $\Gamma (P_{c})$ and $M_{P_c}$ are the width and the mass of the lowest pentaquark from ones  detected by the LHCb  experiment ~\cite{Aaij:2015tga,Aaij:2016phn}.  This state was chosen due to the fact that, most likely, it can have the quantum numbers of the proton.

Thus, the probability of the transition of the proton to the pentaquark is $\varepsilon^2 \lesssim  10^{-5}$. Let us note that this restriction is stronger than the restrictions obtained in most other studies (see, for example, ~\cite{Brodsky:1980pb,Mikhasenko:2012km, Bednyakov:2017vck}).

The authors thank prof. A.K. Likhoded for help and fruitful discussion.
The study is supported by the Russian Found of Basic Research (grant 20-02-00154-A). 
A. Berezhnoy  acknowledges the support from  ``Basis'' Foundation (grant 17-12-244-1).

\bibliographystyle{apsrev4-1}
\bibliography{litr}

\end{document}